\documentclass{article}
\usepackage{spconf,amsmath,graphicx}

\usepackage{enumitem}
\setlist{nosep, leftmargin=14pt}

\usepackage{mwe} 

\usepackage{amssymb}
\usepackage{latexsym}

\usepackage{url}
\usepackage[dvipsnames]{xcolor}
\def\rm{\color{Mahogany}}

\usepackage{hyperref}

\usepackage{subcaption} 
\usepackage{amsmath, amsfonts,graphics} 
\usepackage{graphicx}

\usepackage{tabularx}

\usepackage{svg}

\usepackage{soul}

\title{Multitask Learning for Improved Late Mechanical Activation Detection of Heart From Cine DENSE MRI}
\name{
\parbox{\linewidth}{
\centering
Jiarui Xing$^{ a}$ \qquad
Shuo Wang$^{ d}$ \qquad \\
Kenneth C. Bilchick$^{ d}$ \qquad
Frederick H. Epstein$^{ c}$ \qquad
Amit R. Patel$^{ d}$ \qquad 
Miaomiao Zhang$^{ a, b}$
}}

\address{
\parbox{\linewidth}{
\centering
$^{a}$ Department of Electrical and Computer Engineering, University of Virginia, USA \\
$^{b}$ Department of Computer Science, University of Virginia, USA  \\
$^{c}$ Department of Biomedical Engineering, University of Virginia Health System, USA\\
$^{d}$ School of Medicine, University of Virginia Health System, USA
}}

\usepackage{lipsum}
\begin{document}
%
\maketitle
%
%

\begin{abstract}
The selection of an optimal pacing site, which is ideally scar-free and late activated, is critical to the response of cardiac resynchronization therapy (CRT). Despite the success of current approaches formulating the detection of such late mechanical activation (LMA) regions as a problem of activation time regression, their accuracy remains unsatisfactory, particularly in cases where myocardial scar exists. To address this issue, this paper introduces a multi-task deep learning framework that simultaneously estimates LMA amount and classify the scar-free LMA regions based on cine displacement encoding with stimulated echoes (DENSE) magnetic resonance imaging (MRI). With a newly introduced auxiliary LMA region classification sub-network, our proposed model shows more robustness to the complex pattern cause by myocardial scar, significantly eliminates their negative effects in LMA detection, and in turn improves the performance of scar classification.  To evaluate the effectiveness of our method, we tests our model on real cardiac MR images and compare the predicted LMA with the state-of-the-art approaches. It shows that our approach achieves substantially increased accuracy. In addition, we employ the gradient-weighted class activation mapping (Grad-CAM) to visualize the feature maps learned by all methods. Experimental results suggest that our proposed model better recognizes the LMA region pattern.

\end{abstract}

\section{Introduction}
Cardiac resynchronization therapy (CRT) is widely used to treat cardiac conduction system disorders, such as left bundle branch block and intrinsic myocardial diseases~\cite{abraham2002cardiac,moss2009cardiac}. Standard CRT, however, suffers from a low response rate that is approximately $40\%$~\cite{chung2008results, exner2012contemporary}. It has been shown that the choice of an optimal pacing site has considerable influence on non-response rate, which could be effectively decreased by implanting the CRT left ventricular (LV) lead at areas with late activation~\cite{bilchick2014impact,ramachandran2015singular}. Developing accurate LV activation time measure methods is critical to improve CRT response rate.

A recent work~\cite{auger2017imaging} developed a semi-automatic LV late activation detection method based on the circumferential myocardium strain that is computed from cine DENSE displacement field. The authors first build strain matrices that contain spatial-temporal strain information, and apply active contour algorithm~\cite{kass1988snakes} to estimate the activation time of each location. However, the applicability of this method is limited due to the requirement of a case-by-case model parameter tuning and low optimization efficiency because of an iterative search scheme. To alleviate these problems, a deep learning based late activation prediction model was first introduced in~\cite{cardiacISBI}. Despite the success of achieving significantly higher accuracy and speed, the predicted LMA remains unsatisfactory in challenging cases, i.e., when myocardial scars occur. The existence of scar tissues severely affects and produces activation features that can mislead the LMA detection.

In this paper, we introduce a multi-task learning (MTL) network that simultaneously detects LMA and classify scar-free LMA regions based on cine DENSE MR images. In contrast to the previously proposed deep networks that detect LMA without considering the negative scar effects~\cite{cardiacISBI}, we jointly train an LMA estimation regression network, as well as an auxiliary LMA region classification network that guides the LMA to avoid non-LMA regions. To evaluate the effectiveness of our method, we run tests using real cardiac MR images and compare the predicted LMA regions with the state-of-the-art deep learning-based approaches~\cite{cardiacISBI}. Experimental results show that our approach achieves substantially increased LMA estimation accuracy. Additionally, the attention heatmaps generated by the gradient-weighted class activation mapping (Grad-CAM)~\cite{selvaraju2017grad} indicates that our proposed model focus better on the real LMA features. 
\section{Background}
\textbf{Strain analysis.} The time to the onset of circumferential shortening (TOS) has been proven to have a close linear relationship with electrical activation time and the ability to detect LMA regions~\cite{wyman1999mapping}. Following the same principles, we use the TOS of segmental myocardial regions to measure the cardiac mechanical activation. In order to estimate the TOS, we use circumferential myocardium strain (computed on the displacement field of cine DENSE images)~\cite{budge2012mr} that shows several advantages of: (i) representing myocardial contraction in a relatively low-dimensional space that reduces the cost of network training and risk of overfitting; (ii) robustness to motion artifacts; and (iii) the effectiveness of dyssynchrony quantification and high reproducibility for validation~\cite{budge2012mr}. 

\begin{figure}[h]
    \centering
    \includegraphics[width=0.9\linewidth]{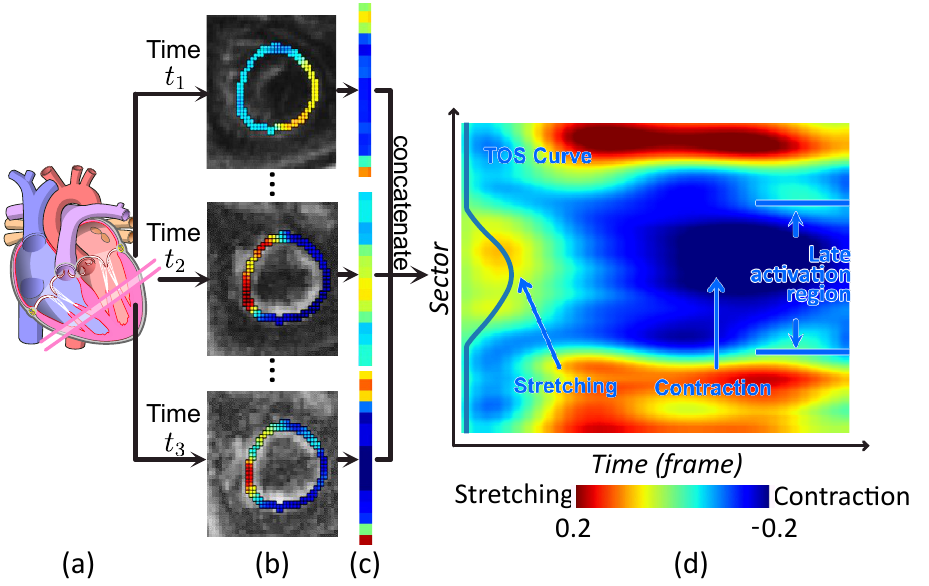}
    \caption{A flowchart of strain analysis of cine DENSE MRIs. Here, $t_1$, $t_2$ and $t_3$ represent different time frames. (a) LV segmental myocardial strains; and (b) generating a strain matrix of all time frames.}
    \label{fig:strain Analysis}
\end{figure}

\textbf{Building Strain Matrix.} Consider a cine DENSE image with $T$ time frames. For each time frame, we compute a $N$-dimensional strain vector from a number of $N$ myocardial sectors, beginning from the the middle of the intersection points of the left and right ventricle and following counter-clockwise order. A $N \times T$ strain matrix that includes information from all time frames is further built by concatenating all strain vectors across time. A curve of TOS values of all myocardial sectors can be overlaid on the strain matrix (see examples in Fig.~\ref{fig:strain Analysis}(d) and (e)).

\textbf{Challenging strain pattern introduced by myocardial scarring.} 
Strain in the scarring region would show stretching in the early frames due to the compression from neighbor sectors, but show no or little contraction in the late frames, which may fool the existing algorithm to wrongly recognize it as LMA. Two strain matrices with TOS curves are shown in Fig.~\ref{fig:strain-mat-with-scar}, where in (a) there is no scar region and the predicted TOS curves from all methods overlap well with the ground truth. In (b), however, both the previously proposed algorithms are misled by the strong early stretching in the scar region and make wrong prediction, while our method gives much more reliable result.

\section{Our Method}

\begin{figure}[h]
    \centering
    \includegraphics[width=0.9\linewidth]{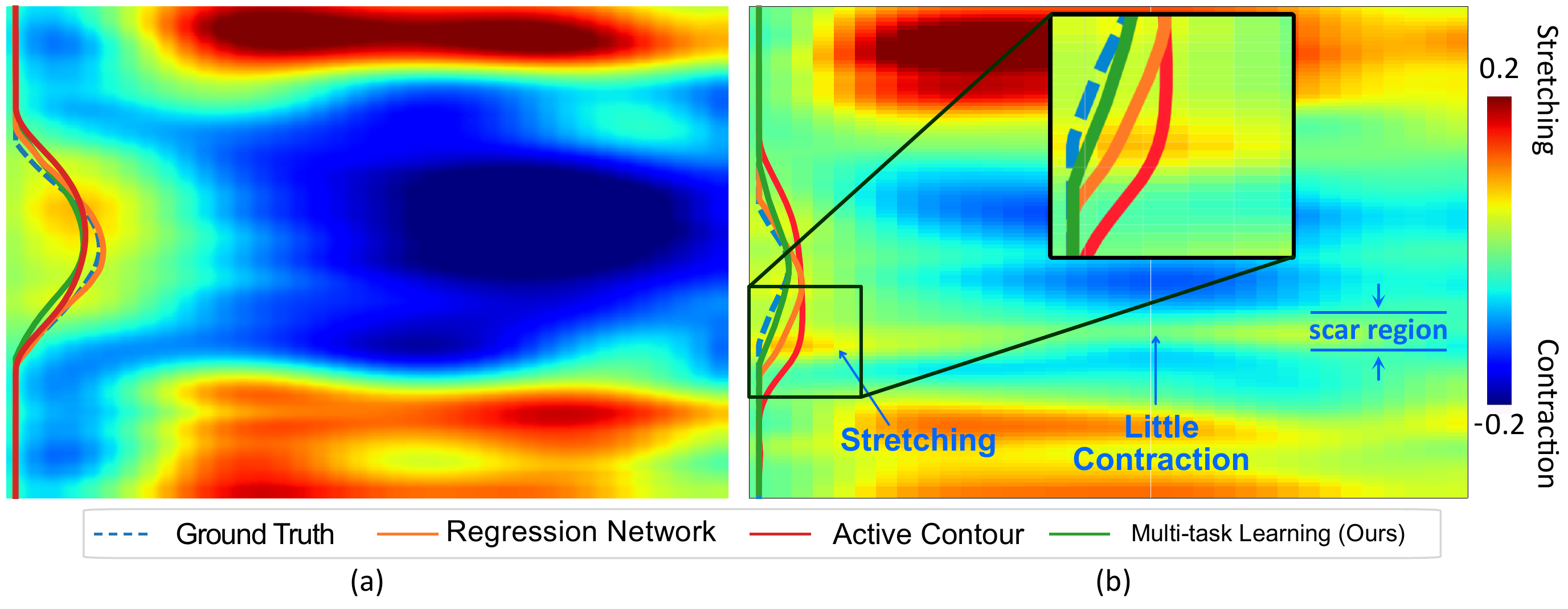}
    \caption{Examples of strain matrices with ground truth and predicted TOS curves.}
    \label{fig:strain-mat-with-scar}
\end{figure}

\textbf{Enhancing algorithm robustness to myocardial scarring by multi-task learning.} To solve the problem of over-estimating the LMA regions, an auxiliary task of determining the LMA region is introduced to solve the problem. When the model is trained only to reduce the difference between ground-truth and predicted TOS curves, the model may focus much more on predicting accurate TOS values than the range of LMA region. For example, in Fig.~\ref{fig:strain-mat-with-scar}(b), while visually the previously proposed models make obvious mistake in the scar region, this error actually contribute little to the total error. On the contrary, the auxiliary LMA region classification task will result in much larger penalty for incorrect LMA region prediction, and thus guide the model to recognize the misleading regions and give more accurate TOS prediction. Besides, adding the auxiliary has been theoretically proven to also adding extra regularization and increase the generalization ability~\cite{baxter1997bayesian,caruana1997multitask}.

\begin{figure*}[!t]
    \centering
    \includegraphics[width=0.85\linewidth]{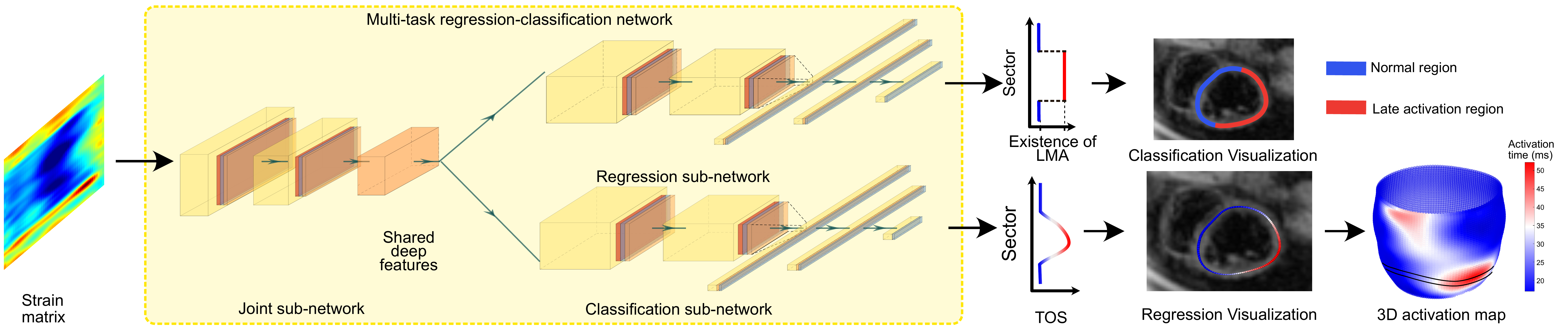}
    \caption{An overview of our proposed multi-task deep learning framework.}
    \label{fig:pipeline}
\end{figure*}

\textbf{Proposed MTL framework.} As is shown in Fig.~\ref{fig:pipeline}, three main components are included in the framework: the joint convolutional sub-network, the regression sub-network, and the classification sub-network. Given the input strain matrix, the joint convolutional sub-network first extracts deep features that are used by both following sub-networks. Taking the deep features as input, the regression sub-network predicts the TOS values of all sectors, while the classification sub-network determines whether each sector has LMA.

The joint convolutional sub-network contains a sequence of convolutional, max-pooling, batch normalization and ReLU layers. Both the classification sub-network and regression sub-network contain the same types of layers as the joint network and multiple fully connected layers. 
A shifted leaky ReLU layer, which is proposed in~\cite{cardiacISBI}, is adopted as the final activation function for the regression network. Compared with other common activation functions, the shifted leaky ReLU layer prevents negative TOS predictions as the TOS value should be always positive, and at the same time avoids the gradient vanishing problem of ReLU~\cite{maas2013rectifier}. 


The loss is the weighted sum of the regression loss, the classification loss and the regularization term:
\begin{align}
    \mathcal{L} 
    &= \mathcal{L}_{reg} + \lambda_{cls}\cdot\mathcal{L}_{cls} + r\cdot \mathcal{R}(w) \nonumber\\
    &= \|\hat{\mathbf{t}} - \mathbf{t}\|_2 + \lambda_{cls}\cdot\text{CE}(\hat{\mathbf{l}}, \mathbf{l}) + r\cdot |\mathbf{w}|, \nonumber
\end{align}

where $\mathbf{t}$ and $\hat{\mathbf{t}}$ are the ground-truth and predicted TOS curves, respectively. $\mathbf{l}$ and $\hat{\mathbf{l}}$ are the ground-truth and predicted one-hot label vectors, which indicate whether each sector has late activation or not. $\mathbf{w}$ is the trainable parameter of the whole network. $|\cdot|$ refers to the $L_1$ norm, and $\mathcal{R}(\mathbf{w})$ works as model regularizer with weighting parameter $r$.  $\text{CE}(\cdot, \cdot)$ denotes the cross entropy, which is defined as $\text{CE}(\hat{\mathbf{l}}, \mathbf{l})
    = - \frac{1}{N_s}\sum_{i}^{N_s}\sum_{j}^{N_c}\hat{\mathbf{l}}_{ij}\log{(\mathbf{l}_{ij})}$.

\textbf{Interpreting the models with Grad-CAM.} To better check whether our model recognizes the scar region, we adapts Gradient-weighted Class Activation Mapping (Grad-CAM) algorithm~\cite{selvaraju2017grad}. Grad-CAM highlights the regions which affects the model prediction most. As a good LMA prediction model should make prediction based on both pre-stretching and late activation patterns, the corresponding Grad-CAM result should highlight both regions. Since the original Grad-CAM works for only a single dimension of the output, we visualize the Grad-CAM of the dimension corresponding to the central sector of the LMA region.

\section{Experiential Results}
\subsection{Experimental Dettings}
\textbf{Data acquisition.} Data were acquired using a 1.5T MR scanner (Avanto, Siemens, Erlangen, Germany) with a four-channel phased-array radiofrequency coil. Cine DENSE was performed in 4 short-axis planes at basal, two mid-ventricular, and apical levels. Cine DENSE parameters included a temporal resolution of 17 ms, pixel size of $2.65 \times 2.65  \text{ mm}^2$ and slice thickness = $8$ mm. Displacement was encoded in two orthogonal directions and a spiral k-space trajectory was used with 6 interleaves per image. Other parameters included: field of view $= 240 \times 240 \text{ mm}^2$, displacement encoding frequency $k_e = 0.1 \text{ cycles/mm}$, flip angle $15^{\circ}$ and echo time $= 1.08 \text{ ms}$.
We have $282$ strain matrices in total from $57$ patients, in which $262$ strain matrices from $53$ patients are used for training, and $20$ matrices from $4$ other patients are used for test. All of the ground-truth TOS curves are manually labeled by experts.

\textbf{\bf Data augmentation and pre-processing.}
Utilizing the circular nature of strain matrices, the data are augmented by shifting the matrices and labels along the sector dimension, which generates new data while keeping the meaningful cardiac strain pattern. Besides, mixup \cite{zhang2017mixup} is also used in our experiments, which generates new data by taking the weighted average of two arbitrary input data and the weighted average of corresponding labels. $3582$ strain matrices and corresponding labels are generated with these methods. Together with the raw training data, in total $3844$ data are used for training. The last $5$ frames of strain matrices are usually noisy and thus trimmed. Moreover, horizontal zero padding and vertical duplication padding are applied to unify the size of strain matrices.

\textbf{Backbone network structure.} In our experiment, the joint sub-network contain 3 successive convolutional layers with the same output channel size as 16 and kernel size as 3. Each the classification and regression sub-network contains 3 convolutional layers, which are followed by fully connected layers with decreasing dimensions. Each of the convolutional layers are followed by a batch normalization layer, a max-pooling layer and a ReLU layer. Each of the fully connected layers are followed by a batch normalization layer and a ReLU layer, except for the last layers, which are followed by the leaky ReLU layers as mentioned in the methodology section.

\textbf{\bf Parameter tuning.} Both the regression and multi-task networks are trained on an Nvidia 2080Ti GPU with $11$ GB RAM over maximum $1000$ epochs using an Adam optimizer~\cite{kingma2014adam} with early stop. The optimal learning rate, batch size, classification loss weight $\lambda_{cls}$ and regularizer weight $r$ are decided through parameter-tuning process using the tree-structured parzen estimator (TPE) \cite{bergstra2011algorithms}. The empirical optimal learning rate, batch size and $r$ for the regression network are $1E-2$, $64$, $10$ and $0.5$, respectively. For the multi-task network, the empirical optimal learning rate, batch size, $\lambda_{cls}$ and $r$ are $1E-3$, $64$, $10$ and $0.1$, respectively. The optimal hyper-parameters for the active contour based method is tuned using Bayesian optimization algorithm \cite{snoek2012practical} on an Intel i5-1035G4 CPU, and the optimal $\lambda=0.60206$, $\beta=4.8778$ and $\gamma=0.43512$.

\subsection{TOS Prediction}
\begin{figure}[!ht]
    \centering
    \includegraphics[width=0.9\linewidth]{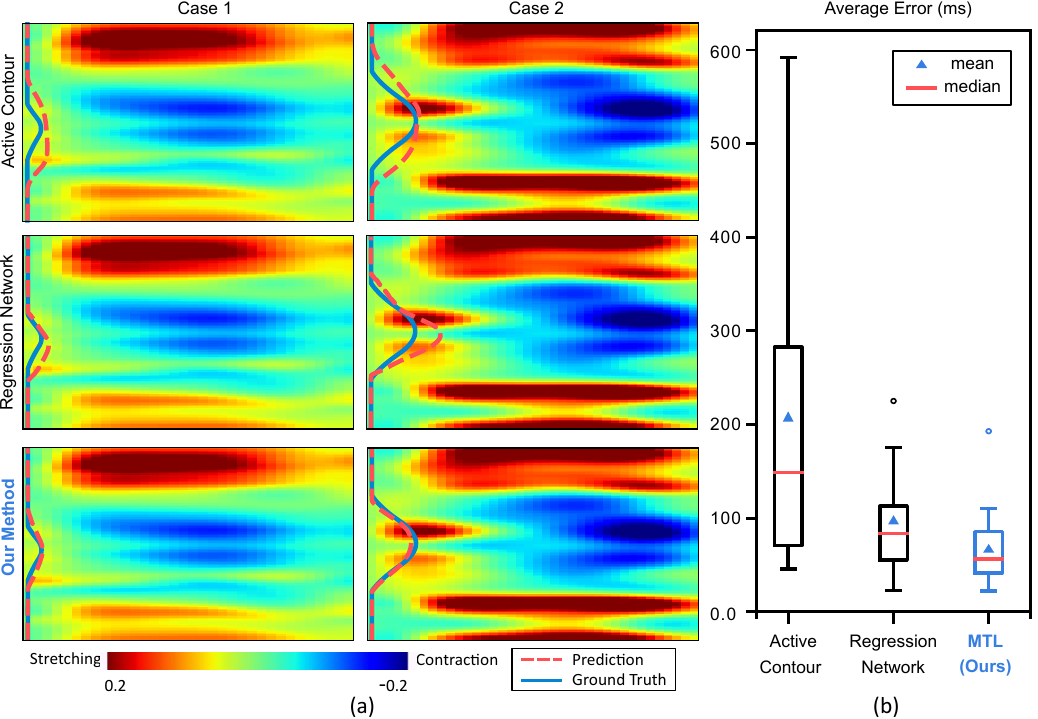}
    \vspace*{-0.05in}
    \caption{Left: Exemplar prediction results of different methods; right: Boxplot of average prediction error on single strain matrix.}
    \label{fig:pred-with-boxplot}
\end{figure}

\textbf{\bf Accuracy.} Three examples of test data prediction of compared methods are shown in Fig.~\ref{fig:pred-with-boxplot}(a). It can be observed that our proposed multi-task network offers the most accurate late activation prediction, in terms of both location and magnitude. The average activation time prediction error on the test dataset of the activation contour-based method, the regression network and our proposed network are $207.5$ms, $97.4$ms and $66.9$ms, respectively, which are also shown in Fig.~\ref{fig:pred-with-boxplot}(b), from which we can see our proposed method offers overall more accurate and robust prediction. Besides, in the case 1 our method successfully recognized the scar region, which is described in the background section, while the other two mistakenly considered the scarring region as late activation region.

\begin{figure}[!h]
    \centering
    \vspace*{-0.05in}
    \includegraphics[width=1.0\linewidth]{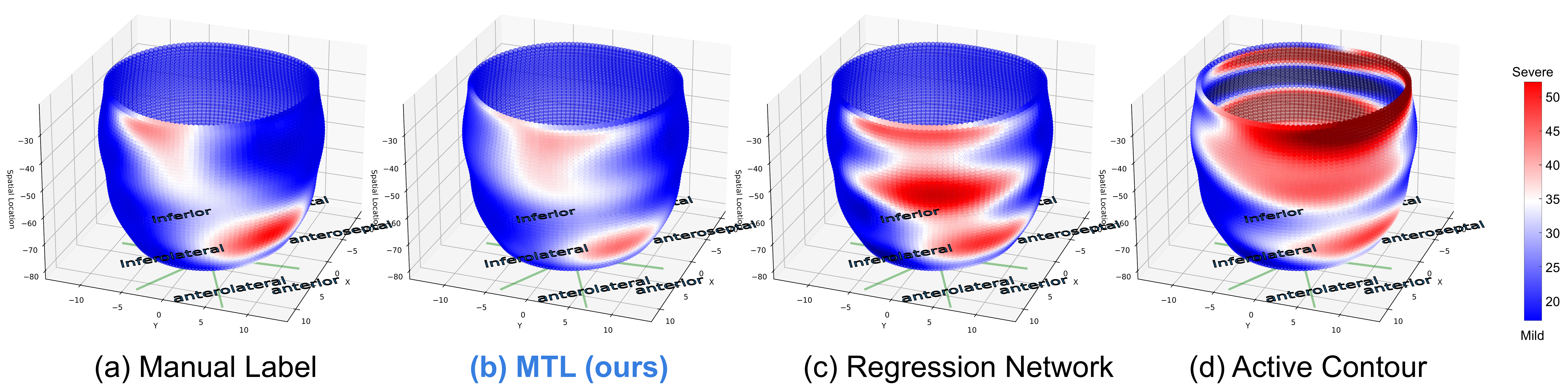}
    \caption{3D Activation Maps built from prediction of different TOS prediction algorithms.}
    \label{fig:3DMaps}
\end{figure}

\textbf{\bf 3D activation maps.} 
We further visualization the distribution of activation time on the myocardium surface as 3D activation maps as are shown in Fig.~\ref{fig:3DMaps}. The regions with large TOS values are shown in red and thus are severe late activating regions, and the regions having small TOS values are shown in blue and are normal regions. Our proposed multi-task network successfully captures both the two major late activation regions at apex inferior and basal inferolateral, and gains significant higher accuracy in both late activation region localization and TOS value prediction.

\begin{figure}[!h]
    \centering
    \includegraphics[width=0.9\linewidth]{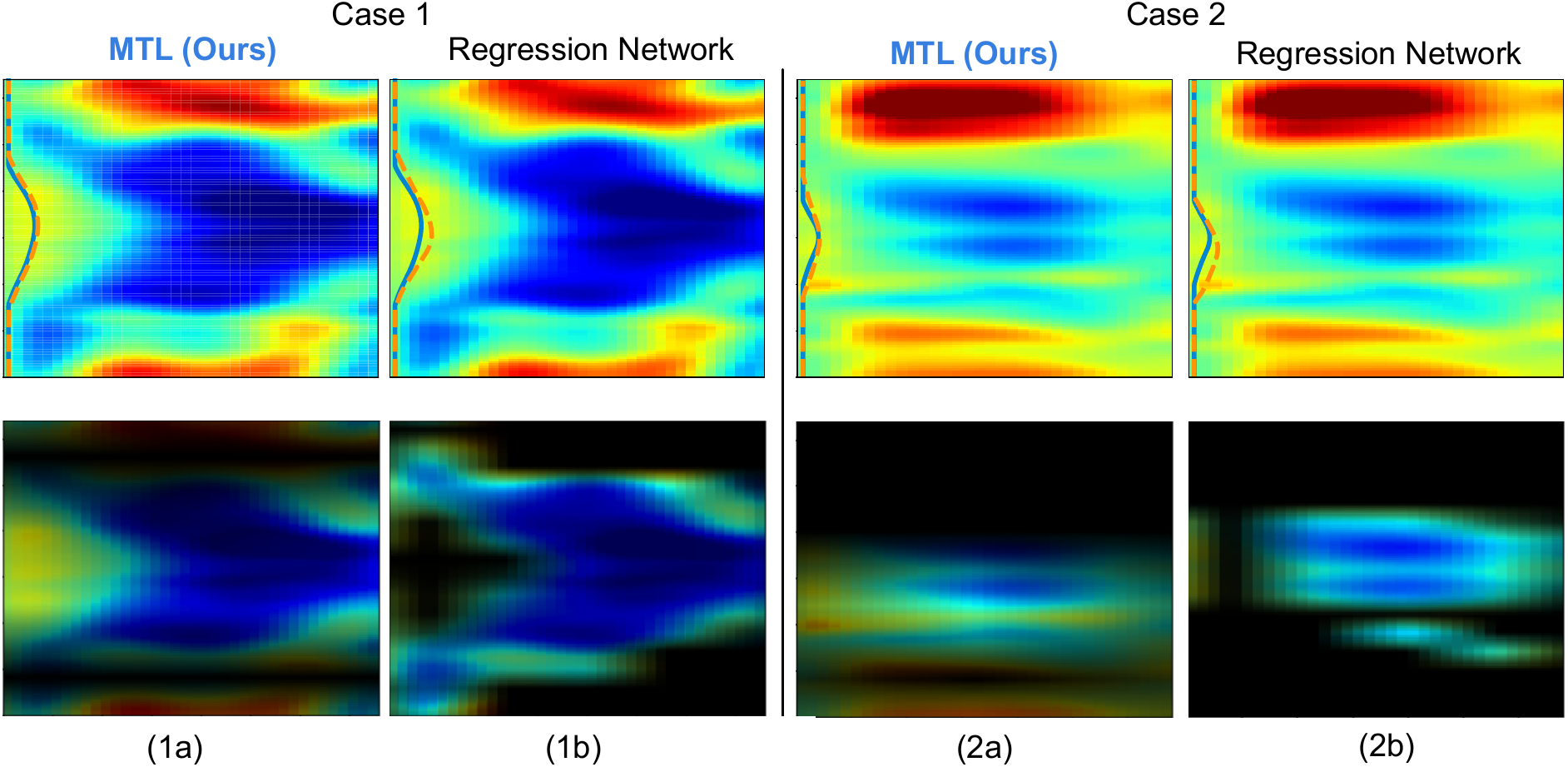}
    \caption{First row: Strain matrices and TOS curves; second row: Strain matrix overlapped with inverse Grad-CAM, where the highlighted regions contribute most to the models' decision.}
    \label{fig:Grad-CAM}
\end{figure}

\textbf{Model interpretation with Grad-CAM.} The Grad-CAM results for the regression model and our MTL model are shown in Fig.~\ref{fig:Grad-CAM}. While we expect the network to recognize the pre-stretching followed by late contraction pattern to identify late activation region, the regression network takes almost only the deep blue late contraction region into consideration and ignores the shallow yellow pre-stretching region when detecting LMA. This explains why it can be misled by the myocardial scarring pattern. On the contrary, the multi-task network focuses on almost all frames near the late activation region, which includes both pre-stretching and late contraction regions.

\section{Discussion}
Our proposed method can be further improved by utilizing additional medical information. For example, we can introduce the scar location information and convert the binary classification task into a multi-class classification task, which detects LMA and scar region in the same time. This will not only further increase the performance on LMA estimation, but also performing valuable scar detection task in the same time.

Meanwhile, our current 3D activation maps of the LV surface is estimated by bilinear interpolation. However, due to the large space between neighbor scanning slices, slight prediction error can cause large different on the 3D activation map. Therefore, we will investigate a better reconstruction to reduce such errors. 

\section{Conclusion}
This paper presented a novel multi-task learning network for an improved cardiac LMA detection. Our newly introduced classification task of identifying non-late-activating regions substantially improves the performance of LMA detection. We evaluated the proposed multi-task network on the cardiac activation time prediction task on CMR images. Experimental results demonstrated that our proposed method outperforms the state-of-the-art~\cite{auger2017imaging,cardiacISBI}. We further examined the LMA feature patterns captured by our model and the baseline algorithms. Results suggested that our method focuses better on the real late activation pattern of pre-stretching followed by contraction. We commit to make all data and code produced in this work publicly available online upon the acceptance of this manuscript.

\bibliographystyle{IEEEbib}
\bibliography{refs}

\end{document}